\documentclass[amsmath,amssymb,nofootinbib,prd]{revtex4}

\usepackage{setspace}
\usepackage{graphicx}

\usepackage{amsmath,amssymb,amsfonts,amsthm}
\usepackage{enumerate}

\begin{document}
\title{Quantum scalar fields interacting with quantum black hole asymptotic regions}

\author{Rodolfo Gambini$^1$,  Jorge Pullin\footnote{Corresponding author. Email: pullin@lsu.edu}$^2$}
\affiliation{1. Instituto de F\'{\i}sica, Facultad de Ciencias, Igu\'a 4225, esq. Mataojo,
11400 Montevideo, Uruguay. \\
2. Department of Physics and Astronomy, Louisiana State University,
Baton Rouge, LA 70803-4001, USA.}

\begin{abstract}
We continue our work on the study of spherically symmetric loop quantum gravity coupled to two spherically symmetric scalar fields, one that acts as a clock. As a consequence of the presence of the latter, we can define a true Hamiltonian for the theory. In previous papers we have studied the theory for large values of the radial coordinate, that is, far away from any black hole or star that may be present. This makes the calculations considerably more tractable. We have shown that in the asymptotic region the theory admits a large family of quantum vacua for quantum matter fields coupled to quantum gravity, as is expected from the well-known results of quantum field theory on classical curved space-time. Here, we study perturbative corrections involving terms that we neglected in our previous work. Using time-dependent perturbation theory, we show that the states that represent different possible vacua are essentially stable. This ensures that one recovers from a totally quantized gravitational theory coupled to matter the standard behavior of a matter quantum field theory plus low probability transitions due to gravity between particles that differ at most by a small amount of energy.

\end{abstract}
\maketitle

\section{Introduction}
Spherically symmetric loop quantum gravity is a good symmetry-reduced laboratory in which one can study black holes, singularity elimination by quantum theory, and other issues, and has been developed over a decade by now \cite{usreview}. However, the introduction of matter has proved problematic. In the vacuum theory one uses a redefinition of the constraints that allows one to turn them into a Lie algebra and complete the Dirac quantization, which at present is not known to exist in the case coupled to matter at the quantum level. Being able to include massless scalar fields is a potentially attractive setting, as it is known to have a rich dynamics that includes black hole formation and also the critical phenomena discovered by Choptuik \cite{choptuik}. 

Here, we would like to expand on our previous papers \cite{previous1,previous2} that considered a spherically symmetric massless scalar field coupled to spherically symmetric gravity in the presence of a clock given by a second scalar field. The latter gives rise to a true Hamiltonian, so one is quantizing a gauge-fixed theory and does not have to worry about constraints. This avenue of using matter clocks in quantum gravity has been considered by other authors as well (see \cite{gravityquantized} for references). In our approach, we exploit the advantages of the simplifications due to spherically symmetric gravity to make progress in defining the relevant quantum operators in a precise way. Our treatment allows us to study a quantum field theory with a natural cutoff provided by the discreteness of quantum gravity. It makes contact with expected results from quantum field theory in a curved space-time. Our framework can in principle accommodate several space-time situations; here we will concentrate on the one that yields quantum field theory on a black hole or other spherical backgrounds. We work in the far asymptotic region, keeping leading terms in the curvature in the calculations.

In this paper, we will consider using approximations to carry out concrete calculations of the space of states of the coupled theory. We will consider the theory for large values of the radial coordinate and expand it in powers of Newton's constant. Since we are in spherical symmetry, that would mean far away from any black hole or star that may be present. This makes the calculations considerably more tractable. In contrast to our previous papers, we will consider terms at the subleading order in the expansion for large distances.  This will allow us to study effects that may arise due to the presence of curvature and how they may modify the usual quantum field theory formulated in a Minkowskian background. We will see that quantum gravity effects add low probability transitions between physical states of the matter field. We concentrate on the low-energy eigenstates of the true Hamiltonian, which correspond to small momentum of the clock and, therefore, lead to small interference of the clock with the system under study. When one gauge fixes using the second scalar field as the clock, the resulting total Hamiltonian is proportional to the momentum of the clock \cite{previous2}. The solutions with low-energy eigenstates approximate those for the theory without being perturbed by the clock scalar field.

The organization of this article is as follows: in the next section we set up the framework, in section 3 we discuss perturbatively the effects of the terms we neglected in our previous papers. We end with a conclusion.

\section{Classical theory: spherical gravity with a scalar field and a  clock}

We consider the Hamiltonian expanded in powers of $G$ (strictly speaking in powers of $G/l_0$) that we introduced in our previous paper \cite{previous2},
\begin{eqnarray}
    H_{\rm true}&=&H_{\rm grav}+H_{\rm matt}=
\frac{    \vert E^\varphi\vert \sqrt{-2 C'}\sqrt{2}}
{4 \sqrt{\pi G}\,l_0^2}\nonumber\\
&&
-\frac{
\vert E^\varphi\vert 
\sqrt{2\pi G}\sqrt{-2 C'}
\left(
2 \left(\phi'\right)^2 x^4-8 E^\varphi \phi' K_\varphi P_\phi x-4\left(E^\varphi\right)^2 \rho_{\rm vac}+2 P_\phi^2
\right)}
{
4 \left(E^\varphi\right)^2 (-C') l_0^2
}.\label{31}
\end{eqnarray}

The phase space of the theory is that of vacuum spherically symmetric gravity, consisting of the radial triad $E^\varphi$, its conjugate momentum $K_\varphi$ (the radial triad and its conjugate momentum have been gauge fixed), and the scalar field $\phi$ and its conjugate momentum. The scalar field of the clock and its conjugate momentum have also been gauge fixed and therefore do not appear. As discussed in our previous paper \cite{previous2}, $C$ is the Hamiltonian constraint of vacuum gravity, $\rho_{\rm vac}$ is the expectation value of the scalar field energy in the vacuum. The subtraction of this term allows to assume that the matter term does not pertub the gravitational one which allows us to treat the matter term as a perturbation. $l_0$ is a spatial length that appears in the gauge fixing and can be physically interpreted as characterizing the space-time domain in which the scalar field behaves well as a clock.

As is commonly done in spherically symmetric loop quantum gravity one takes a kinematical basis of quantum eigenstates of the operators $\hat{E}^x$ and $\hat{E}^\varphi$ obtained by the direct product of a one-dimensional loop representation along a graph in the radial direction times a Bohr compactification in the transverse direction. That is,
\begin{eqnarray}
\hat{E}^x_j \vert k_1,\ldots, k_N,\mu_1,\ldots, \mu_n\rangle&=&k_j \ell_{\rm Planck}^2\vert k_1,\ldots, k_n,\mu_1,\ldots, \mu_n\rangle,\\
\hat{E}^\varphi_j \vert k_1,\ldots, k_n,\mu_1,\ldots,\mu_n\rangle&=&\mu_j \ell_{\rm Planck}\vert k_1,\ldots, k_N,\mu_1,\ldots, \mu_n\rangle.
\end{eqnarray}
In terms of these, the discrete version of $C$, the vacuum Hamiltonian constraint, is,
\begin{equation}
    \hat{C}_j=-j \Delta \left(1-2 \Lambda +\frac{\sin\left(\rho \hat{K}_{\varphi,j} \right)^2}{\rho^2} -\frac{\left(\hat{E}^x_{j+1}-\hat{E}^x_j\right)^2}{4 \Delta^2 \left(\hat{E}^\varphi_j\right)^2}\right) + 2GM,
\end{equation}
where $\rho$ is the polymerization parameter of the Bohr compactification (not to be confused with the quantity $\rho_{\rm vac}$ that appears later on, which as discussed in our previous paper \cite{previous2} is the energy of the vacuum that leads in spherical symmetry to a solid angle defect $\Lambda =2\pi G\rho_{\rm vac}$).  The operator $\hat{E}^x_j$ commutes with $\hat{H}_{\rm true}$ and therefore is a constant of the motion that in the spin network representation has eigenvalues $k_j\ell_{\rm Planck}^2$ with $k_j$ integers. In order to simplify things, we choose an equally spaced lattice with $E^x(x)=x^2$,  $x_j=j \Delta+x_0 $, with $j>0$, $\Delta=n \ell_{\rm Planck}$ the lattice spacing and with this choice $k_j={x_j}^2/\ell_{\rm Planck}^2$, $n$ a small positive integer and as we mentioned $x_0\gg r_S=2GM$ so we are in the asymptotic region.

At a quantum level, the purely gravitational part is given by,
\begin{equation}
    \hat{H}_{\rm grav} 
    \Psi\left(l_1,\ldots, l_N\right)=
    \sum_j 
    \frac{x_j F  \sqrt{2\Delta}\sqrt{-\delta l_j}}
    {\sqrt{\frac{l_j-r_S}{x_j}+1 -2\Lambda}}
    \Psi\left(l_1,\ldots, l_N\right),
\end{equation}
with $F=\left(2 \sqrt{2\pi}\ell_{\rm planck}l_0^2\right)^{-1}$ and $\delta l_j=l_{j+1}-l_j$. Here, $l_j$ are the eigenvalues of $\hat{C}_j$.

  In our previous paper we studied the Hamiltonian and its properties in the region $x\gg r_S$ with $r_S$ the Schwarzschild radius at the zeroth order \cite{previous2}. For this, we considered normalizable states that approximated well the states of the continuous spectrum of the Hamiltonian. Here, we will find it more convenient to work directly in the improper eigenstates of the Hamiltonian.  We will use time-dependent perturbation theory to study the corrections to the Hamiltonian that we considered in our previous paper. We will use improper eigenstates because the spectrum of the Hamiltonian is continuous, and in perturbation theory, one usually uses the eigenstates.

We will divide the matter part into a zeroth order term in the expansion in $1/x$ that corresponds to the $H_{\rm matt}$ from the previous paper \cite{previous2} and a first order term that corresponds to the asymptotic corrections that we ignored in that paper.

Let us consider the zeroth order portion of the matter Hamiltonian,
\begin{equation} \hat{H}_{\rm matt}^{(0)}=
    \frac{\sqrt{2\pi}\sqrt{G}}{\sqrt{-2C'(x)^{(0)}}l_0^2 E^\varphi(x)^{(0)}}
\left(\hat{\phi}'(x)^2 x^4- 2 \left(E^\varphi(x)^{(0)}\right)^2 \rho_{\rm vac} + \hat{P}(x)^2\right),
\end{equation}
where from now on we call the momentum of the scalar field $P$ instead of $P_\phi$ to simplify notation.
This Hamiltonian is obtained by taking expectation values on the gravitational variables with the normalizable gravitational state considered in our previous paper, leading to the following expectation values for the gravitational variables,
\begin{eqnarray}
E^\varphi(x)^{(0)} &=& \frac{x}{\sqrt{1-2\Lambda}},\\
K_\varphi(x)^{(0)} &=& 0,\\
C'(x)^{(0)}&=& \frac{ x^2}{l_0^2 \pi^2}.
\end{eqnarray}

The above Hamiltonian can be rewritten as,
\begin{equation} \hat{H}_{\rm matt}^{(0)}(x)=
\frac{ 2\pi^{3/2}}{ l_0} \sqrt{G}\sqrt{1-2\Lambda}\left(\frac{\hat{\phi}'(x)^2 x_k^2}{2}+\frac{\hat{P}(x)^2}{2 x^2}-\frac{\rho_{\rm vac}}{2(1-2\Lambda)}\right),
\end{equation}
where as we discussed in our previous papers, $\rho_{\rm vac}$ is a counterterm of the energy of the vacuum that we absorb in the solid deficit angle $\Lambda=2\pi G\rho_{\rm vac}$. The quantity $l_0$ is used in the definition of the clock $\varphi=t/l_0^2$ with $\varphi$ the scalar field used as a clock (in our previous paper \cite{previous2} we called $\psi$ the scalar field and $\phi$ the clock one) and $t$ the asymptotic time. The physical interpretation of $l_0$ is the range of validity of the clock, which determines the size of the asymptotic region that we can analyze with it. By range of validity, we mean a region where there is a non-vanishing clock scalar field and its momentum is small.

We recognize the standard scalar field Hamiltonian on the lattice in the above expression (up to the constant term proportional $\rho_{\rm vac}$, which will be evaluated later),
\begin{equation}
\hat{H}_{{\rm matt},j} = \sum_j \left(\frac{\hat{P}_j^2}{2 x_j^2 \Delta}+\frac{\left(\hat{\phi}_{j+1}-\hat{\phi}_j\right)^2 x_j^2}{2 \Delta}\right).\label{11}
\end{equation}

and the total Hamiltonian (gravity plus matter), at zeroth order, is,
\begin{equation}
    \hat{H}^{(0)}_j = \frac{x_j^{3/2} \sqrt{-\delta l_j \Delta}}
    {2 \sqrt{\pi}\ell_{\rm Planck}
    l_0^2\sqrt{l_j-r_S+\left(1-2\Lambda\right)x_j}}-\frac{2\pi^{3/2}\sqrt{1-2\Lambda}{\sqrt{G}}}{l_0} \hat{H}_{\rm matt}.
\end{equation}

 We consider the elements of a continuous basis for the gravitational part of the Hamiltonian,
\begin{equation}
   \Psi\left(\vec{l}\right)= \prod_j \langle l_j \vert f^{(0)}_j+\epsilon f^{(1)}_j \rangle=\prod_j \delta(-l_j+f^{(0)}_j+\epsilon f^{(1)}_j)\label{2},
\end{equation}
with $f^{(1)}\ll x_0$ and $f^{(0)}$ chosen to recover the states that lead to the matter Hamiltonian discussed in the previous paper. $\epsilon$ is a small quantity to emphasize that the term in $f^{(1)}$ has a small contribution and the limit $\epsilon\to0$ corresponds to the results of our previous paper. Here, $\delta$ is the Dirac delta. In order to compute the first-order correction to the Hamiltonian, we shall expand in $\epsilon$ and evaluate the first-order coefficient in $\epsilon$.  In the basis (\ref{2}) we choose $f^{(0)}$ to yield $\left(C^{(0)}(x)\right)'$, that is, the Hamiltonian of vacuum gravity that makes $H_{\rm matter}$ take the Minkowskian form. In our previous paper \cite{previous2} we carried out a similar construction for normalizable states. The choice that leads to this result is,
\begin{equation}
    f^{(0)}_{j+1}=-\frac{x_j^2\Delta}{l_0^2\pi^2}+f^{(0)}_j,
\end{equation}
that is,
\begin{equation}
    f^{(0)}_{j+1}=\frac{\left( \Delta j+x_0\right)^2\Delta}{l_0^2 \pi^2}+f^{(0)}_j.
\end{equation}

The recursion relation can be solved as,
\begin{equation}
     f^{(0)}_{j+1}=-
   \frac{\left(\left(j^{2}+\frac{3}{2} j +\frac{1}{2}\right) \Delta^{2}+3 \mathit{x_0} \Delta  \left(j +1\right)+3 \mathit{x_0}^{2}\right) j \Delta}{3 \mathit{l_0}^{2} \pi^{2}},
\end{equation}
which satisfies,
\begin{equation}
    \hat{C}'_{{\rm tot},j}\Psi\left(\vec{l}\right)\equiv \frac{\hat{C}_{j+1}-{C}_j}{\Delta}\Psi\left(\vec{l}\right)=\left[\frac{ \epsilon}{\Delta}\left(f^{(1)}_j- f^{(1)}_{j+1}\right)+\frac{ x_j^2}{l_0^2\pi^2}\right]\Psi\left(\vec{l}\right).\label{17}
\end{equation}

The solutions of the continuous spectrum of the gravitational part are for $f^{(1)}\ll f^{(0)}$ and take the form,
\begin{equation}
\Psi_{\rm gr} =\prod_j \delta\left(l_{j+1}+\frac{\left(j^{2}+\frac{3}{2} j +\frac{1}{2}\right) \Delta^{2}+3 \mathit{x_0} \Delta  \left(j +1\right)+3 \mathit{x_0}^{2}} {j \Delta}
    -\epsilon f^{(1)}_{j+1}\right).
\end{equation}

The matrix elements of $\hat{E}^\varphi$ in the improper basis are,
\begin{equation}
    \langle \vec{l}^1\vert \hat{E}^\varphi_j\vert \vec{l}^2\rangle = \frac{x_j \delta\left(l^1_j-l^2_j\right)}{\sqrt{\frac{l^1_j-r_S}{x_j}+1-2\Lambda}},
\end{equation}
from where we can read the form of the multiplicative operator to first order in $(r_s-l_j^1)/x_j$, and recalling that the $l_j$ are eigenvalues of the $\hat{C}_j$,
\begin{equation}
\left(\hat{E}^\varphi_j\right)_{(1)} = \frac{r_S-\hat{C}_j}{2\left(1-2\Lambda\right)^{3/2}},
\end{equation}
and as before,
\begin{equation}
\left(\hat{E}^\varphi_j\right)_{(0)} = \frac{x_j}{\sqrt{1-2\Lambda}}.
\end{equation}
For the inverses, we have,
\begin{equation}
\left(\hat{E}^\varphi_j\right)^{-1}_{(1)}=\frac{\hat{C}_j-r_S}{2 \sqrt{1-2\Lambda}x_j^2},
\end{equation}
\begin{equation}
\left(\hat{E}^\varphi_j\right)^{-1}_{(0)}=\frac{\sqrt{1-2\Lambda}}{x_j}.    \end{equation}

The above operators are diagonal in the improper basis. For the connection, it is a bit more complicated. We start by defining the basis of eigenstates of $\hat{E}^\varphi$, 
\begin{equation}
    \hat{E}^\varphi_j \vert \vec{\mu}\rangle = \mu_j  \ell_{\rm Planck} \vert \vec{\mu}\rangle,
\end{equation}
and given the eigenbasis for $\hat{C}$ we already considered $\vert \vec{l}\rangle$ we compute,
\begin{equation}
    \langle \vec{l}'\,\vert \hat{K}_{\varphi,j}\vert \vec{l}\,\rangle =
    \int \langle \vec{l}'\,\vert \vec{\mu}\rangle \langle \vec{\mu}\,\vert \hat{K}_{\varphi,j}\vert \vec{l}\,\rangle d\vec{\mu}= 
    \int \langle \vec{l}'\,\vert \vec{\mu}\rangle {\rm i}\frac{d}{d\mu_j}\langle \vec{\mu}\,\vert \vec{l}\,\rangle d\vec{\mu}.
\end{equation}
The above eigenstates are direct products of the eigenstates at each site 
\begin{equation}
    \vert \vec{l}\,\rangle =\prod_j \vert l_j\,\rangle,
\end{equation}
and similarly for $\vert \vec{\mu}\rangle$. We also have that (see our previous paper),
\begin{equation}
\langle \mu_j\,\vert l_k\rangle=
\frac{\sqrt{2}}{2 \sqrt{\ell_{\rm Planck}}}\delta\left(
\mu_j -\frac{x_j}{\ell_{\rm Planck} \sqrt{\frac{l_j-r_S}{x_j} +1 -2\Lambda}}
\right)   
\left(
\frac{l_j-r_S}{x_j}+1 -2\Lambda
\right)^{-3/4}\delta_{jk}
\end{equation}
and therefore,
\begin{equation}
    \langle l_i \vert \hat{K}_\varphi\vert l'_j\rangle = \ell_{\rm Planck}\left[2
    \left(\frac{l_j -r_S}{x_j}+1-2\Lambda\right)\delta'\left(l_i-l_j'\right)+\frac{3}{2 x_j} \delta(l_i-l_j')\right]\sqrt{1-2\Lambda +\frac{l_j-r_S}{x_j}}.
\end{equation}

\section{Perturbative analysis for the first order correction to the asymptotic approximation}

Taking into account that $\hat{E}^\varphi_j$ and $\hat{C}'_j$ have zeroth- and first-order terms, and including $\hat{K}_{\varphi,j}$ as a first-order correction and that $\hat{C}'_j=\frac{\delta \hat{l}_{j}}{\Delta}$, 
the total Hamiltonian expanded to second power in $\epsilon$ is,
\begin{equation} \hat{H}^{(0)}=\sum_j\left\{
\frac{x_{j}  \sqrt{{| {\delta \hat{l}_{j}}{\Delta}|}}}
{2 \sqrt{\pi}\, \mathit{\ell_{\rm Planck}} \,\mathit{l_0}^{2} \sqrt{\frac{{\hat{l}_{j}-\mathit{r_S}}}{x_j}+1-2 \Lambda}}-\left(
\frac{\, x_{j}^{2}  \sqrt{G}\, \left(\hat{\phi}_{j+1}-\hat{\phi}_j \right)^{2}}{\mathit{\Delta l_0}}
+\frac{ \sqrt{G}\, \hat{P}_j^{2} }{x_{j}^2  \Delta \mathit{l_0}}\right)\pi^{3/2}\sqrt{1-2 \Lambda},\right\}
\end{equation}
\begin{eqnarray}
\hat{H}^{(1)}&=&\sum_j\left\{
\frac{\sqrt{1-2 \Lambda}\,  \pi^{\frac{7}{2}} \sqrt{G}\, 
\left(\hat{\phi}_{i+1}-\hat{\phi}_i\right)^{2} {{\hat{C}^{(1)'}}}_j \mathit{l_0}}
{4 \Delta}
+\frac{\left(-\hat{l}_{j}+\mathit{r_S} \right) x_{j} \pi^{\frac{3}{2}} \sqrt{G}\, \left(\hat{\phi}_{i+1}-\hat{\phi}_i\right)^{2}}{2l_0 \Delta\sqrt{1-2 \Lambda}\, }\nonumber\right.\\
&&
\left.
+\frac{4 \pi^{\frac{3}{2}} \sqrt{G}\, \mathit{\hat{K}_{\varphi,j}^{(1)}}\left[\left(\hat{\phi}_{j+1}-\hat{\phi}_j\right), \mathit{\hat P}_j\right]_{+}  }{\Delta\mathit{l_0}x_j}
+\frac{\pi^{\frac{7}{2}} \sqrt{G}\, \hat{P}_j^{2} \sqrt{1-2 \Lambda}\, {{\hat{C}^{(1)'}}}_j \mathit{l_0}}{4 x_{j}^{4}\Delta }+\frac{\pi^{\frac{3}{2}} \sqrt{G}\, \hat{P}_j^2 \left(-\hat{l}_{j}+\mathit{r_S} \right)}{2 \sqrt{1-2 \Lambda}\, x_{j}^{3}\Delta l_0 } \right\}
\end{eqnarray}
where $[\;,\;]_{+}$ is the anticommutator. The operator $\hat{C}^{(1)}$, when acting on $\psi\left(\vec{l}\right)$, is the term in (\ref{17}) proportional to $\epsilon$ and $\hat{C}^{(0)}$ yields the term independent of $\epsilon$. 

The energy at zeroth order (of the gravitational field) per site for improper states $\delta\left(l_j-f^{(0)}_j-f^{(1)}_j\right)$ is,
\begin{equation}
    E^{(0)}_j=\frac{x_j \sqrt{\left\vert \Delta \left(\frac{\Delta x_j^2}{l_0^2\pi^2}-f^{(1)}_{j+1}+f^{(1)}_j\right)\right\vert}}{2 \sqrt{\pi} \ell_{\rm Planck} l_0^2 \sqrt{1-2\Lambda -\frac{f^{(0)}_j+f^{(1)}_j-r_S}{x_j}}}.
\end{equation}

Using time dependent perturbation theory, as for instance discussed in \cite{messiah}, extended to the case of improper states, we have schematically that the transition probability (densities) between  (improper) eigenstates of the non-perturbed Hamiltonian $H^{(0)}$, which we call for simplicity $a$ and $b$, is given by,
\begin{equation}
    W_{a\to b} = \vert H^{(1)}_{ab}\vert^2 \left(2 \frac{\sin^2 \omega_{ab}t }{\omega_{ab}^2}\right)\label{32},
\end{equation}
with 
\begin{equation}
H^{(1)}_{ba}=\langle b\vert H^{(1)}\vert a\rangle,
\end{equation}
and, 
\begin{equation}
    \omega_{ab} = {E^0_a -E^0_b},\label{34}
\end{equation}
and we recall we are working with $\hbar=c=1$.

The goal is to compute the probability densities for the situation we are considering and to analyze their consequences. 
To apply perturbation theory, we need the eigenstates of the zeroth order Hamiltonian. In our previous paper, we have analyzed the gravitational part of this Hamiltonian. We need to consider the matter part. Neglecting the point polymerization of the scalar field, it turns out that the resulting Hamiltonian on a spin network is the same as that of a scalar field on a lattice, as we discussed in (\ref{11}).

We need to expand the scalar field present in that expression in creation and annihilation operators,
\begin{equation}
    \hat{\phi}_{v,j}
= \left(\hat{a}_v \exp(-i\omega_v t)+\hat{a}^\dagger_v \exp(i\omega_v t)\right) \frac{\sin(k_v x_j)}{\sqrt{\pi v} x_j},\label{36}
 \end{equation}
where $k_v =2\pi v/(j_N \Delta)$
with $j_N=(l_0-x_0)/\Delta\sim l_0/\Delta$ is the number of nodes in the asymptotic region where we are studying the field that ranges in $x$ from $x_0$ to $l_0$. The integer $v$ characterizes the different modes of the field. We recognize, at zeroth order, the usual form for the Hamiltonian of a scalar field,
\begin{equation}
\hat{H}_{{\rm matt}} =\sum_v k_v 
\frac{\hat{a}_v \hat{a}^\dagger_v
+\hat{a}^\dagger_v \hat{a}_v}{2}+O(\Delta).
\end{equation}
As a consequence, the eigenstates of the complete Hamiltonian are given by,
\begin{equation}
    \Psi^{(0)}_{\rm total} =\prod_j \delta\left(-l_j+f^{(0)}_j+f^{(1)}_j\right)\Phi_{k_{v_1},\ldots, k_{v_q}},\label{38}
\end{equation}
with $\Phi$ is the eigenstate of the zeroth order Hamiltonian of the scalar field that has the modes $k_{v_1}$ to $k_{v_q}$ excited. Since the spin network introduces a natural cut-off, the scalar field has a discrete spectrum. If the state of the matter part remains invariant,  (\ref{34}) takes the form,
\begin{equation}
   \omega_{ff'}= {E^{(0)}_f-E^{(0)}_{f'}}={\sqrt{\Delta}} \sum_j x_j \frac{ \sqrt{-2 \delta {f^{(1)}}'_j l_0^2 \pi^2 +\Delta x_j^2}-\sqrt{-2\delta f^{(1)}_j l_0^2 \pi^2 +\Delta x_j^2}}{2 \pi^{3/2} \ell_{\rm Planck} l_0^3 \sqrt{1-2\Lambda -\frac{f^{(0)}_j}{x_j}}} .
\end{equation}

The expectation value in the gravitational part of the order one  Hamiltonian, neglecting terms of order $\ell_{\rm Planck}^2$, between states like those in (\ref{38}) with $f^{(1)}$ and ${f'}^{(1)}$, is given by
\begin{equation}
\langle \hat{H}^{(1)}_j\rangle_{\rm grav} =
\frac{\pi^{\frac{3}{2}} \left(2 \left(\mathit{f^{(1)}}_{j +1}-\mathit{f^{(1)}}_{j}\right) \pi^{2} \left(-\frac{1}{2}+\Lambda \right) \mathit{l_0}^{2}+\left(-\mathit{f^{(0)}}_{j}-\mathit{f^{(1)}}_{j}+\mathit{r_S} \right) x_{j} \ell_{\rm Planck} \right) }{\sqrt{1-2 \Lambda}\, \mathit{l_0} x_{j}^{2}}
\delta\left(f^{(1)}_j-f'{}^{(1)}_j\right)
\mathit{\hat{H}_{{\rm matt},j}}.\label{124}
\end{equation}

In order to manage this expression, it is good to give a concrete form for the ${f^{(1)}}_j$'s involved. We choose
$f^{(1)}_j=f^{(1)}$, a constant. 

 We will compute transition probabilities for different values of the constant to get an idea of how they behave. 

With this choice, since $x_j$ will be typically large, the form of the expectation value in the gravitational part of the first order  Hamiltonian yields an operator acting on the matter variables,
\begin{equation}
    \langle \hat{H}^{(1)}_j\rangle_{\rm grav} =
    \frac{\ell_{\rm Planck} x_j^2}{2\sqrt{\pi}\sqrt{1-2\Lambda}l_0^3}
    \delta\left(f^{(1)}_j-f'{}^{(1)}_j\right)
    \mathit{\hat{H}_{{\rm matt},j}}.
\end{equation}
The presence of the $x_j^2$ factor on the right-hand side modifies the weights of the different terms of the order zero Hamiltonian and will imply the existence of transitions between states of the scalar field due to coupling to gravity.

To simplify the calculations, we will go to the continuum limit, but we keep an ultraviolet cut-off in the momentum variable ${2\pi}/{\Delta}$. This is an excellent approximation, given that the spin network sites in spherical symmetry can be made as close as $\ell_{\rm Planck}^2/r_S$ due to the condition of quantization of the areas of symmetry. We choose it to be proportional to $\ell_{\rm Planck}$ to have a uniform lattice. In that limit, the equivalent expression to (\ref{36}) is,
\begin{equation}
    \hat{\phi}(x,t)= \int dk \left(\hat{a}_k e^{-i k t}+a^\dagger_k e^{i k t} \right) \frac{\sin(kx)}{\sqrt{\pi k}},
\end{equation}
and similarly for the field momentum,
 \begin{equation}
    \hat{P}(x,t)= i\int dk \left(-k \hat{a}_k e^{-i k t}+k a^\dagger_k e^{i k t} \right) \frac{\sin(kx)}{\sqrt{\pi k}}.
\end{equation}
Substituting these expressions into the first-order Hamiltonian, we get
\begin{eqnarray}
    \langle \hat{H}^{(1)}(x)\rangle_{\rm grav} &=&
    \left(6 \pi^{3/2} \sqrt{1-2\Lambda}\sqrt{k k'}l_0^3\right)^{-1}
    \left(
    \hat{a}_k \hat{a}^\dagger_{k'} {\mathrm e}^{\mathrm{-i} t \left(k -{k'} \right)}
    +\hat{a}^\dagger_{k} \hat{a}_{{k'}} {\mathrm e}^{\mathrm{i} t \left(k -{k'} \right)}
    +\hat{a}_k \hat{a}_{{k'}} {\mathrm e}^{\mathrm{-i} t \left(k 
    +{k'} \right)}+\hat{a}^\dagger_{k} \hat{a}^\dagger_{{k'}} {\mathrm e}^{\mathrm{i} t \left(k +{k'} \right)}\right)\nonumber\\
    &&\times\left(\cos((k'-k) x) x^2 k' k
    -x  \sin((k-k') x) k'
    +\sin(k x)\sin(k' x)\right).
\end{eqnarray}
Integrating in $x$, and recalling that $k_n=2\pi n/(l_0-x_0)$ with $n$ an integer,
\begin{eqnarray}
    \langle \hat{H}^{(1)}\rangle_{\rm grav}^{k\ne k'} &=&
\frac{\ell_{\rm Planck}}{ \pi^{\frac{3}{2}} \sqrt{1-2 \Lambda}\, \mathit{l_0}^{3}}\left[
\frac{2 \mathit{(l_0-x_0)} \sqrt{k \mathit{k'}}\, }{\left(k -\mathit{k'}\right)^2}
\right]\nonumber\\
&&\times
\left(\hat{a}_{k} \hat{a}^\dagger_{\mathit{k'}} {\mathrm e}^{\mathrm{-i} t \left(k -\mathit{k'} \right)}+\hat{a}^\dagger_{k} \hat{a}_{\mathit{k'}} {\mathrm e}^{\mathrm{i} t \left(k -\mathit{k'} \right)}+\hat{a}_{k} \hat{a}_{\mathit{k'}} {\mathrm e}^{\mathrm{-i} t \left(k +\mathit{k'} \right)}+\hat{a}^\dagger_{\mathit{k'}} \hat{a}^\dagger_{k} {\mathrm e}^{\mathrm{i} t \left(k +\mathit{k'} \right)}\right), \label{44}
\end{eqnarray}
\begin{eqnarray}
    \langle \hat{H}^{(1)}\rangle_{\rm grav}^{k =k'} &=&
\frac{\ell_{\rm Planck}}{ \pi^{\frac{3}{2}} \sqrt{1-2 \Lambda}\, \mathit{l_0}^{3}}\left[
\left(\mathit{x_0} k \,\mathit{(l_0-x_0)}^{2}+\mathit{x_0}^{2} k \mathit{(l_0-x_0)} +\frac{k \,\mathit{(l_0-x_0)}^{3}}{3}+\frac{\mathit{(l_0-x_0)}}{k}\right) \right]\nonumber\\
&&\times
\left(\hat{a}_{k} \hat{a}^\dagger_{\mathit{k'}} {\mathrm e}^{\mathrm{-i} t \left(k -\mathit{k'} \right)}+\hat{a}^\dagger_{k} \hat{a}_{\mathit{k'}} {\mathrm e}^{\mathrm{i} t \left(k -\mathit{k'} \right)}+\hat{a}_{k} \hat{a}_{\mathit{k'}} {\mathrm e}^{\mathrm{-i} t \left(k +\mathit{k'} \right)}+\hat{a}^\dagger_{\mathit{k'}} \hat{a}^\dagger_{k} {\mathrm e}^{\mathrm{i} t \left(k +\mathit{k'} \right)}\right) \delta_{k,k'},
\end{eqnarray}
and we have recalled  for the last integration the original form of the discrete $k$ and $k'$ in the spin network, which leads to the Kronecker instead of Dirac deltas. 

We therefore see that the perturbative Hamiltonian can create and annihilate pairs of matter particles, such as the last terms in (\ref{44}). However, such terms do not conserve energy (in the sense of the matter portion of $H^{(0)}$) and are therefore heavily suppressed in (\ref{32}). 
There are energy conserving contributions to second order, but in the asymptotic regions these terms are negligible.  This ensures that in the asymptotically flat limit one recovers the usual quantum field theory treatment, in which there is no particle production from the geometry. However, in other background geometries this could lead to particle production, hinting at the emergence of Hawking radiation. This could lead to effects of interest, for instance, in cosmological backgrounds or closer to the horizon.

\section{Conclusions}

We have studied spherically symmetric gravity coupled to a spherical scalar field using a second spherical scalar field as a clock. We concentrate in the asymptotic region, keeping terms that we had neglected in previous publications. The effect of the terms is to induce transitions in the states of the scalar field due to interactions with gravity. Transitions are of low probability even when they do not induce changes in the energy of the scalar field. We therefore see the emergence of quantum gravity effects in the asymptotic region, but they are small, as expected. It would be more interesting to get closer to the horizon, where one could study quantum gravity corrections to Hawking radiation due to the back reaction. There are several methods proposed that would allow us to deal with that region perturbatively \cite{various}. Summarizing, we see the emergence of quantum field theory on a quantum spacetime with spherical symmetry in the far asymptotic region that includes effects involving gravity that start to depart from usual quantum field theory on curved spacetime.

\section{Acknowledgements}
We thank Richard Price for comments.
This work was supported in part by grant  NSF-PHY-2206557, funds of the
Hearne Institute for Theoretical Physics, CCT-LSU, Fondo Clemente Estable
FCE 1 2019 1 155865.

\end{document}